# Title: Variability in Fermi–Pasta–Ulam–Tsingou Arrays Prevents Recurrences

**Authors:** Heather Nelson[1], Mason A. Porter[2], Bhaskar Choubey[1].

**Affiliations:**

[1]Department of Engineering Science, University of Oxford, Parks Road, Oxford OX1 3PJ, UK.

[2]Department of Mathematics, University of California, Los Angeles, 520 Portola Plaza, Los Angeles, CA 90095, USA.

**Abstract:** In 1955, Fermi, Pasta, Ulam, and Tsingou reported recurrence over time of energy between modes in a one-dimensional array of nonlinear oscillators. Subsequently, there have been myriad numerical experiments using homogenous FPUT arrays, which consist of chains of ideal, nonlinearly-coupled oscillators. However, inherent variations — e.g., due to manufacturing tolerance — introduce heterogeneity into the parameters of any physical system. We demonstrate that such tolerances degrade the observance of recurrence, often leading to complete loss in moderately sized arrays. We numerically simulate heterogeneous FPUT systems to investigate the effects of tolerances on dynamics. Our results illustrate that tolerances in real nonlinear oscillator arrays may limit the applicability of results from numerical experiments on them to physical systems, unless appropriate heterogeneities are taken into account.

**Main Text:**

Fermi, Pasta, Ulam, and Tsingou undertook what is widely believed to be the first nonlinear computational experiments, and they presented their now-famous results in a May 1955 technical report (1). They excited a one-dimensional (1D) array of linear oscillators coupled by nonlinear springs (see Fig. 1A) in the first mode, and they expected that the system's nonlinearity would lead to equipartition of energy between all modes. However, to their surprise, their numerical experiments instead showed that although energy did start to transfer from the first mode to higher modes, over time it actually appears to return to the first mode. This phenomenon constitutes a "recurrence" of energy between modes. Such FPUT recurrences have been the subject of much research in the past half century, and the FPUT numerical experiments have led to a wealth of work on recurrences, nonlinear lattice systems, and other prominent topics (2–9).

The equation of motion for a quadratically-coupled 1D FPUT lattice (i.e., the FPUT-$\alpha$ model) is

$$\ddot{x}_i = x_{i+1} + x_{i-1} - 2x_i + \alpha\left[(x_{i+1} - x_i)^2 - (x_i - x_{i-1})^2\right], \qquad (1)$$

where the index *i* denotes the *i*th oscillator. In Fig. 1B, we show the first four energy modes for a 1D array of $N = 64$ oscillators with a nonlinear coupling strength $\alpha$ of 0.25, which is similar to the original report from FPUT.

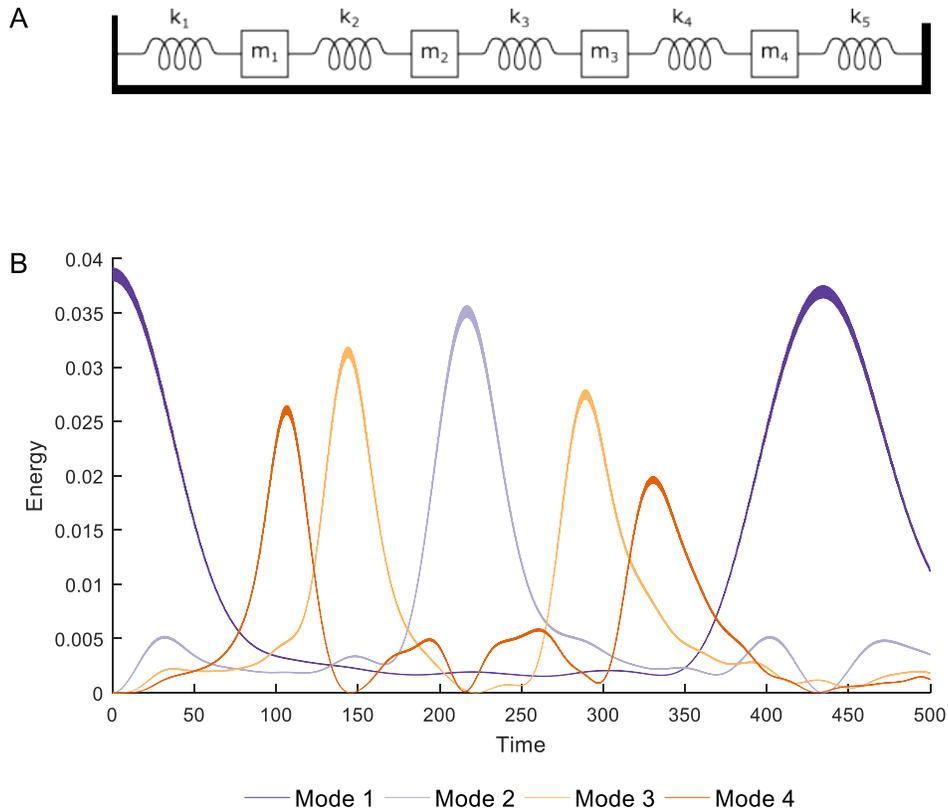

**Fig. 1.** (A) A system of masses coupled via nonlinear springs in a similar manner to that used by Fermi, Pasta, Ulam, and Tsingou in their numerical experiments. (B) FPUT recurrences in a chain of 64 oscillators (see Eqn. 1) with a coupling strength of $\alpha = 0.25$.

  The recurrence phenomenon in these simulations has significant implications in nonlinear dynamics, and extensive computational and theoretical work has examined FPUT recurrences *(2, 5, 8, 10-11)*. However, there has been very little experimental evidence of such recurrences *(12-13)*. We propose that this paucity of evidence is due to variability in system components due to manufacturing tolerances. Most analytical and computational studies of FPUT system have assumed that the component oscillators are homogeneous *(5, 6, 10, 16–18)*.

  It is typically simpler to study homogeneous lattice systems than heterogeneous ones, but it is difficult (if not impossible) to manufacture a large array of identical oscillators. Inherent manufacturing variations always introduce a tolerance, as is observed in any mechanical or electronic system *(19–26)*. Hence, the individual oscillators of any real array will not be identical.

  Additionally, most physical parameters associated with real devices also vary with environmental conditions such as temperature, pressure, and humidity. Therefore, even if one could build an array of identical oscillators, it is very likely that the individual oscillators would change while the system is operating. Moreover, individual oscillators in a system would likely change in different ways, leading to further variability between the oscillators.

Although such tolerances are well-studied in many physical systems *(27, 28)*, we are not aware of any investigations on their effects on FPUT lattices. Impurities and defects in lattices have attracted attention, including work on topics like Anderson phenomena and defect modes, but this has an extremely different focus from our engineering-oriented work *(9, 14)*. Additionally, it is not clear how to extrapolate the results of these studies to phenomena such as tolerance-based inhomogeneities in arrays.

To experimentally observe recurrences in practical lattice systems, it is important as a model scenario to understand the effects of tolerance on an FPUT system. This will enable quantification of the constraints under which one can built such a system, if indeed it is possible. To do this, we study the FPUT system under the influence of typical manufacturing tolerances of physical systems.

**Method**

In an FPUT lattice (see Eqn. 1), each oscillator has its own linear stiffness, and the oscillators interact with each other through nonlinear coupling. A simple mass–spring system, such as the one shown in Fig. 1A, has tolerances as in the following equation:

$$\ddot{x}_i = t_{i+1}x_{i+1} + t_{i-1}x_{i-1} - 2t_i x_i + \alpha\left[(t_{i+1}x_{i+1} - t_i x_i)^2 - (t_i x_i - t_{i-1}x_{i-1})^2\right], \quad (2)$$

where $t_i$ represents the individual tolerance of element $i$.

Typical passive electronic components have tolerances of ±0.1%, ±1%, ±5%, and ±10% *(29)*. We study the system in Eqn. 2 to investigate the impact of such variability on FPUT recurrences.

In this paper, we consider the most basic scenario that was described in the original FPUT report *(1)*. We focus on a 1D FPUT-$\alpha$ array with an initial condition in which all of the energy is in the first mode. Unless we state otherwise, all simulations have 64 oscillators and a nonlinear coupling strength of $\alpha = 0.25$. For each oscillator, the initial condition is

$$x_i = \sin\left(\frac{\pi i}{N+1}\right), \quad (3)$$

$$\dot{x}_i = 0, \quad (4)$$

where $N$ is the number of oscillators.

The code that we use to simulate the FPUT equations is based on that published by Dauxois et al. (2005), and it uses a standard method for solving such equations *(32)*. To ensure that our results are not the product of the chosen software, we use both MATLAB (with a standard Runge–Kutta algorithm in its ODE45 solver) and Python (with the SCIPY ODE solver that implements a standard LSODA algorithm). We randomly generate the tolerance values for each simulation from a Gaussian distribution. We use this choice, because it is the most common one in manufacturing *(33–35)*. For a tolerance of $\tau$ %, this entails a value of $t_i$ that we draw from a Gaussian with a mean of 1 and standard deviation of $\sigma = 1/3 \times 0.01\tau$. (For example, for 1% tolerance, the standard deviation is 0.0033.) With a 6$\sigma$ width (±3$\sigma$), we note that 99.73% of the values of $t_i$ fit within the interval $[1 - 0.01\ \tau, 1 + 0.01\ \tau]$. We floor any outlying values of $t_i$ to be on the corresponding edge of this interval.

In light of manufacturing tolerances, the exact value of each parameter is difficult to determine. Therefore, to help understand the effects of a random tolerance spread, we use Monte Carlo simulations and study 100 different sets of tolerance values applied to the FPUT system, while keeping fixed all other system parameters and the initial condition. We run multiple simulations that each use an independent draw of the tolerance values, and then examine the dynamics of the oscillator arrays. The figures in this article are representative of the majority of our simulation results. In Fig. 2, we show example results for a tolerance of ±1%, and we note that Fig. 2A is representative of the response in more than 85% of the cases. However, for some combinations of tolerance values, recurrence is occasionally weak (Fig. 2B) or breaks down (Fig. 2C).

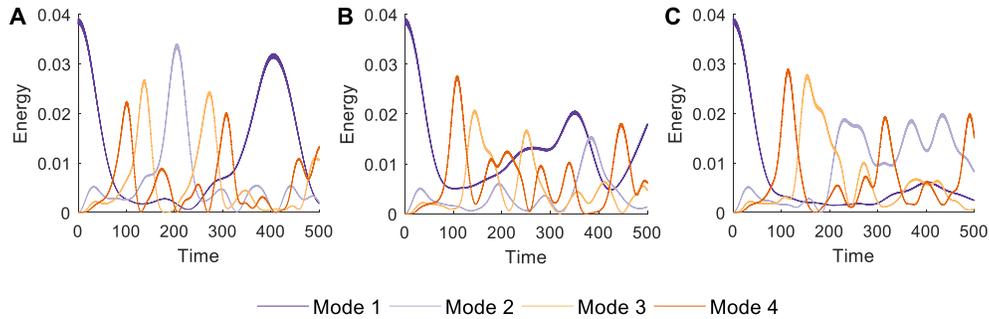

**Fig. 2.** Simulations of a 1D FPUT-$\alpha$ array with 64 oscillators and a coupling strength of $\alpha = 0.25$ showing examples of recurrence at a tolerance of ±1%.

**Effect of tolerance on recurrence**

With a tolerance of ±0.1%, we observe that the FPUT system in Eqn. 2 usually exhibits a similar recurrence to that of the ideal system in Eqn. 1 (see Fig. S1). However, the addition of tolerance does introduce a slight change in the recurrence timescale. In this and subsequent figures, we show a typical result from a Monte Carlo simulation with one randomly selected set of values of $t_i$ from the distribution.

On increasing the tolerance to ±1%, the recurrence phenomenon starts to differ considerably from the ideal scenario of homogeneous system components (see Fig. 3A). Increasing the tolerance further to ±5% (see Fig. 3B), we see even less energy transfer to the higher modes, and very little recurrence of energy appears to be taking place. At ±10% (see Fig. 3C), this steady decline in the quality of the recurrence is even more prominent.

These results are noteworthy. Any real oscillatory system has mismatches between the components. We have shown that with extremely tight tolerances of ±0.1%, the output is close to the ideal, in the sense that the FPUT recurrences resemble those from the case of homogeneous oscillators. However, even at tolerances as small as ±5%, the recurrence phenomenon departs substantially from that in the ideal scenario. To produce a system with a tolerance of ±1%, one requires a high degree of precision in manufacture, which is expensive and often impractical.

Additionally, due to accumulation of tolerances, individual oscillators in electronic and mechanical systems often have stacked tolerances of 5–10% even when they are constructed of individual components with very tight tolerances. At these values, recurrence starts to break down. This suggests that with typical mechanical and electronic systems, one may never see FPUT recurrences at all.

**Significance of tolerances on linear versus nonlinear terms**

The classical FPUT system has equations with two parts: a discretized linear diffusion term and a nonlinear coupling term. To identify whether tolerances in the linear or nonlinear terms contribute more to the breakdown of recurrence, we independently apply tolerances to the linear and nonlinear terms. As the nonlinear coupling leads to recurrence, our initial expectation was that any tolerance in nonlinear coupling parameters should lead to a more severe breakdown of recurrence than incorporating tolerance in the linear parts. However, our observations suggest that this is not the case.

We show our results of incorporating tolerance only in the linear parts of an FPUT-$\alpha$ array in Figs. 3D–F. One can observe that the system is not exhibiting full recurrence, though some partial recurrence does seem to be occurring. As we show in Figs. 3G–I, incorporating tolerance in only the nonlinear terms has, for a fixed amount of tolerance, a comparable effect to incorporating it in only linear terms. When we incorporate tolerance in both the linear and nonlinear terms in an FPUT-$\alpha$ array, we observe more energy transfer into the lower modes than in the above two scenarios (see Figs. 3A–C). This is particularly evident at larger tolerance values.

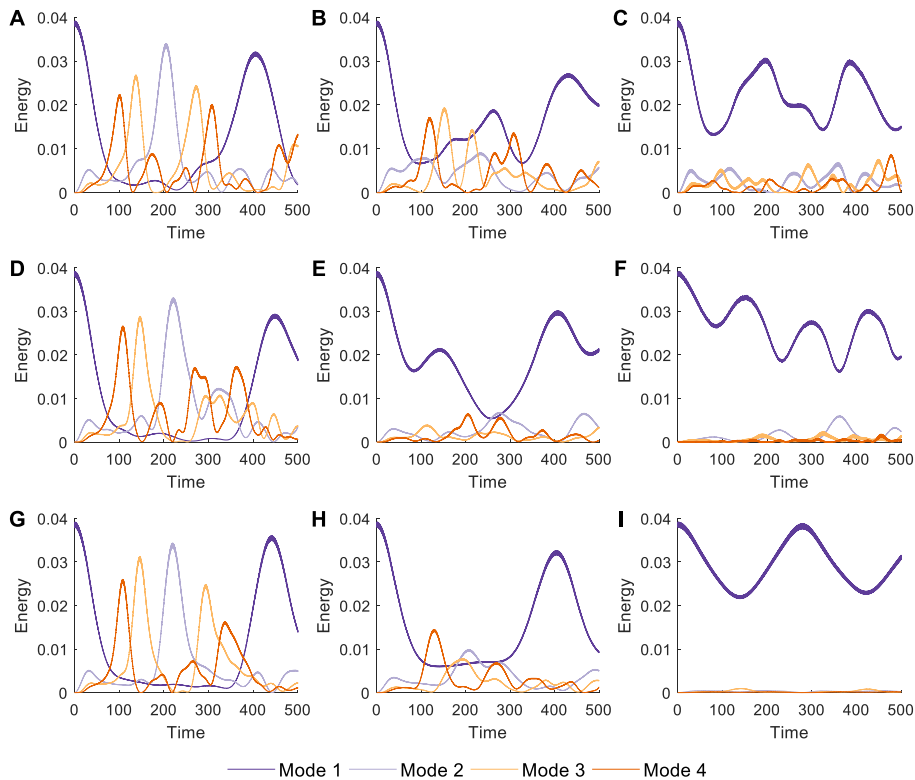

**Fig. 3.** The effect of tolerance on different parts of a 1D FPUT-$\alpha$ array with a coupling strength of $\alpha = 0.25$ and N = 64 oscillators.
Tolerance in both linear and nonlinear terms: (A) ±1%, (B) ±5%, and (C) ±10%.
Tolerance in the linear terms: (D) ±1%, (E) ±5%, and (F) ±10%.
Tolerance in the nonlinear terms: (G) ±1%, (H) ±5%, and (I) ±10%.

**Impact of tolerance on arrays with different numbers of oscillators**

As we have seen, recurrence breaks down as tolerance increases in a 1D FPUT array of 64 oscillators. It is worth examining what happens in arrays with different numbers of oscillators. In our exploration, we compare the results for 1D arrays with 8, 16, 32, 64, and 128 oscillators using a coupling strength of $\alpha = 0.25$ and a tolerance of ±5%.

In Figure 4, we illustrate that 1D arrays with a larger number of oscillators experience a larger impact from incorporating tolerances. One can see clearly that recurrence is strong with only 8 oscillators (see Fig. 4A), whereas recurrence has broken down completely when there are 128 oscillators (see Fig. 4E). This means that in light of typical manufacturing tolerances, one can only safely observe recurrence in an array of very small size. As the size increasers, the recurrence stops occurring at even smaller tolerances and we may never be able to see tolerances in any large sized array.

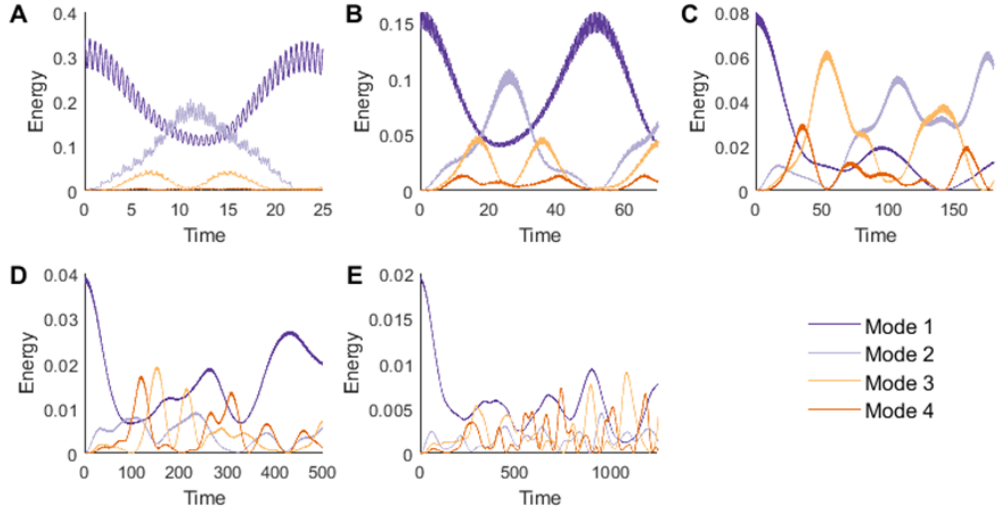

**Fig. 4.** Simulations of 1D FPUT-$\alpha$ arrays with different numbers of oscillators at a tolerance of ±5% with a coupling strength of $\alpha = 0.25$. (A) 8 oscillators, (B) 16 oscillators, (C) 32 oscillators, (D) 64 oscillators, and (E) 128 oscillators. [Panel (D) appeared previously in Fig. 3.]

**Asymmetric Coupling**

In our previous simulations, we showed results from symmetrically-coupled FPUT systems. However, in a variety of physical realizations of such a system, including any electronic ones, forward and backward coupling is not always the same. This asymmetry leads to another source of error, which we incorporate into the following asymmetric FPUT-$\alpha$ array:

$$\ddot{x}_i = t_{i+1}x_{i+1} + t_{i-1}x_{i-1} - 2t_i x_i + \alpha\left[t^a_{i}(x_{i+1} - x_i)^2 - t^a_{i-1}(x_i - x_{i-1})^2\right], \quad (5)$$

where $t_i$ and $t^a_i$ represent the tolerances of oscillator $i$. We conduct simulations of the FPUT system in Eqn. 5 and show our results in Fig. 5, Fig. 6 and Fig. S1.

Comparing the results of our asymmetric system to those of our symmetric system in Eqn. 2, we observe that the impact of tolerance is comparable in the two systems when we incorporate tolerance in both the linear and nonlinear terms. However, recurrence in the asymmetric FPUT system breaks down for a smaller tolerance than it does in the symmetric FPUT system. Additionally, as we increase the number of oscillators in a system, we observe that recurrence breaks down for a smaller number of oscillators in the asymmetric FPUT system than in the symmetric system.

One interesting observation is that incorporating tolerance in only the nonlinear terms of the symmetric FPUT system has a comparable effect to that of adding it to only the linear terms, whereas incorporating tolerance in only the nonlinear terms in the asymmetric FPUT system seems to have significantly less impact than we expected, as we observe recurrence that is close to that of an ideal asymmetric system.

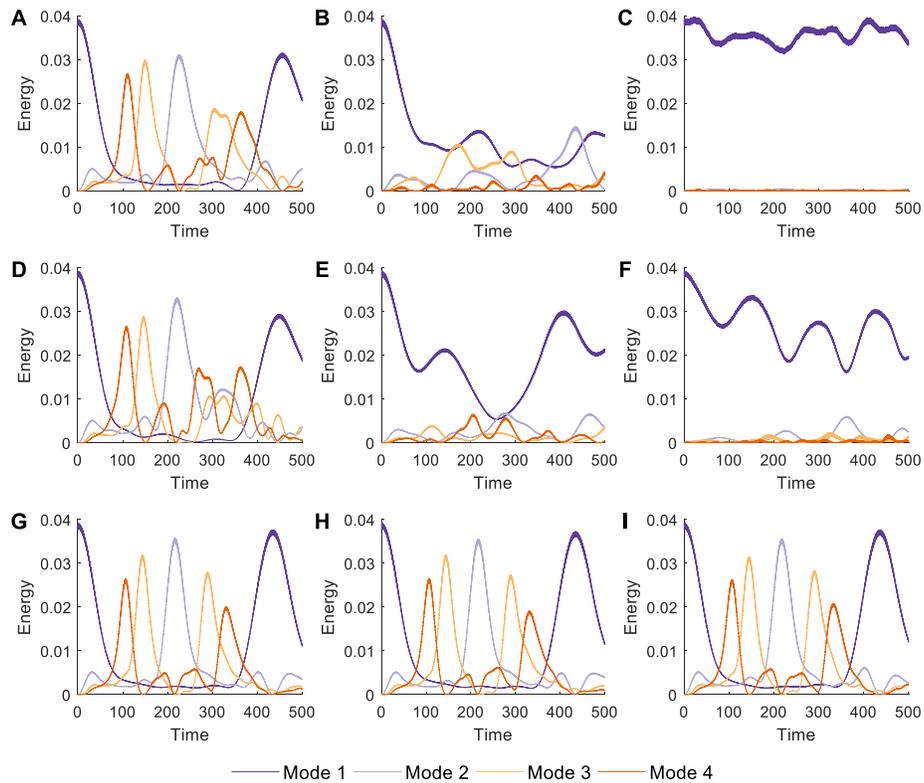

**Fig. 5.** The effect of tolerance on different parts of an asymmetric 1D FPUT-$\alpha$ array with a coupling strength of $\alpha = 0.25$ and N = 64 oscillators.
Tolerance in both linear and nonlinear terms: (A) ±1%, (B) ±5%, and (C) ±10%.
Tolerance in the linear terms: (D) ±1%, (E) ±5%, and (F) ±10%.
Tolerance in the nonlinear terms: (G) ±1%, (H) ±5%, and (I) ±10%.

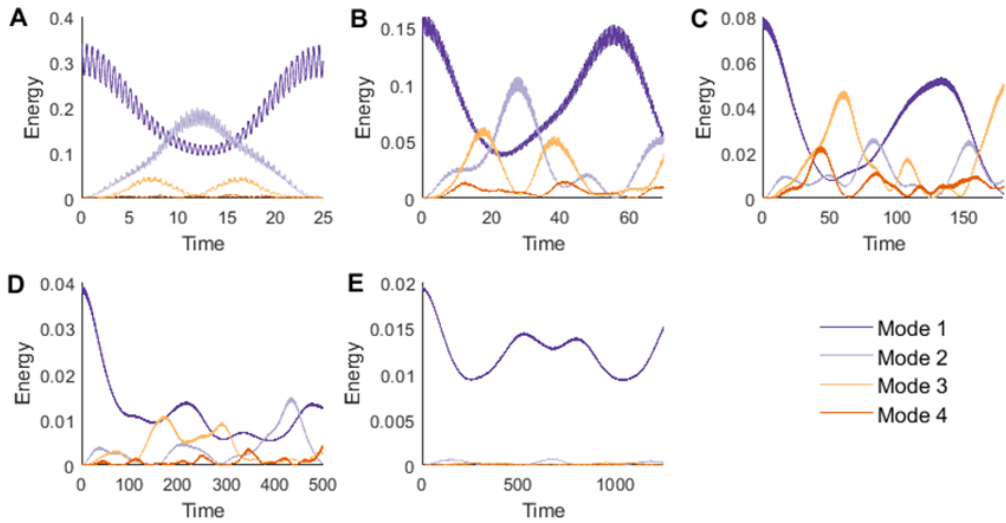

**Fig. 6.** Simulations of asymmetric 1D FPUT-$\alpha$ arrays with different numbers of oscillators at a tolerance of ±5% with a coupling strength of $\alpha = 0.25$. (A) 8 oscillators, (B) 16 oscillators, (C) 32 oscillators, (D) 64 oscillators, and (E) 128 oscillators. [Panel (D) appeared previously in Fig. 5.]

**Conclusions and Discussion**

The numerical experiments reported by Fermi, Pasta, Ulam, and Tsingou in 1955 have led to an extensive body of computational, theoretical, and experimental work in nonlinear systems. However, it is commonly assumed in studies of FPUT and other lattice systems that the units are homogeneous. Physical systems, in contrast, are heterogeneous by nature, and this can affect recurrence phenomena.

In the present paper, we examined the effect of incorporating heterogeneity on a 1D FPUT array. Such heterogeneity arises from the inherent tolerances of various manufacturing processes, so one must take it into consideration in laboratory experiments.

The results of our simulations illustrate that tolerance has a significant impact on recurrences in an FPUT system, destroying it in a 64-element system for tolerance values that lie within the typical range for manufacturing tolerances. However, by reducing the number of oscillators in an FPUT system, one retains recurrence for some reasonable amount of tolerance before it breaks down.

By controlling manufacturing tolerance between nominally identical components, it may be possible to observe recurrence in an FPUT system with a small number of oscillators. However, tight tolerances are hard to achieve and are often very expensive. Therefore, producing such a system may not be practical. In large arrays of oscillators, even very small tolerances will surely eliminate any chance of seeing recurrence in practice.

In this article, we used a typical sinusoidal initial condition in a 1D FPUT array; we have not yet explored other initial conditions. There are numerous studies that investigate FPUT systems with other initial conditions, alongside other considerations (such as driven or damped systems *(5, 17)*). One can also examine similar considerations in other lattice systems, such as granular crystals, which have Hertzian interactions between components *(30, 31)*. Incorporating tolerances into parameters for other initial conditions (and other variants of FPUT systems and other types of nonlinear lattices) can lead to behaviors other than those that we have discussed in this article. Because tolerance has such a major impact on a standard FPUT system with sinusoidal input, it would be unreasonable to assume that other scenarios will be affected only minimally. Consequently, investigations of FPUT systems and nonlinear lattices need to be revisited with tolerance in mind.

There has also been work in other related areas from which it is desirable to draw inspiration for additional work. For example, in studies of synchronization on networks, there is also a long history of examining the collective properties of coupled phase oscillators (a rather different type of system from the one that we study) with natural frequencies drawn from some distribution *(15)*. With practical laboratory experiments in mind, it is crucial to conduct systematic investigations of incorporating tolerance into those and other systems.

**Acknowledgments:** We thank Alejandro Martínez for helpful discussions. **Funding:** HN acknowledges support from the EPSRC under its DTA scheme.

**Author contributions:** Bhaskar Choubey conceptualized the problem of hardware validation of FPUT while Heather Nelson conceptualized variability in these arrays. Heather Nelson undertook the code writing, data generation and its analysis. Resources, supervision, funding acquisition and further data analysis was provided by Bhaskar Choubey. Mason Porter visualized the data and helped validate it. Paper writing was undertaken by all three authors, with initial draft by Heather Nelson.

**Competing interests:** None declared.

**Materials & Correspondence:**

The data and the code is available from the authors upon reasonable request. Material request should be send to Heather Nelson and other correspondence to Bhaskar Choubey.

**Supplementary Materials:**

Figures S1–S2

# Supplementary Materials for

Variability in Fermi–Pasta–Ulam–Tsingou Arrays Prevents Recurrences

Heather Nelson, Mason A. Porter, Bhaskar Choubey.

Correspondence to: heather.nelson@eng.ox.ac.uk

**This PDF file includes:**

Figs. S1 to S2

**Supplementary Text**

Results for 1D FPUT-α Arrays with ±0.1% tolerance

    Figures S1 and S2 show the results of the numerical experiments for a symmetric and an asymmetric FPUT system with tolerance of ±0.1%.

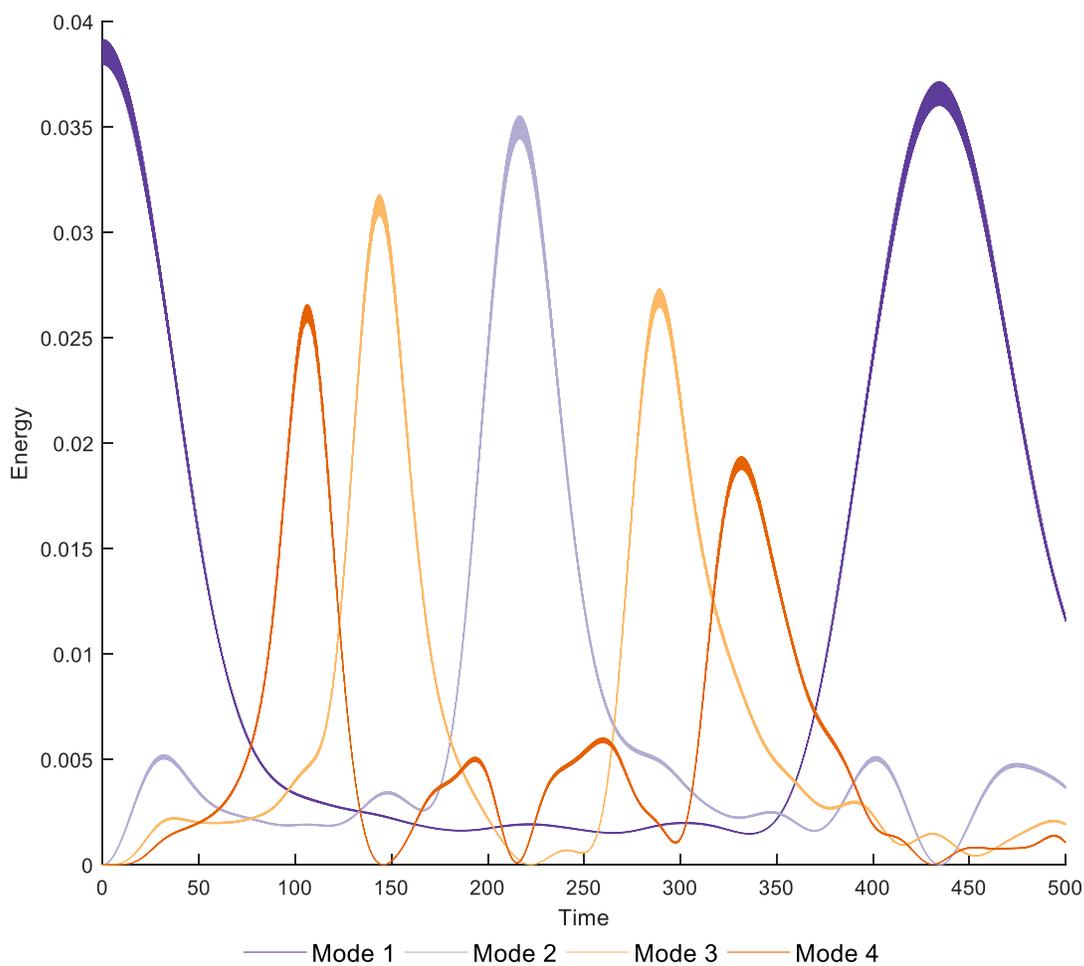

**Fig. S1.**
Simulations of a 1D FPUT-$\alpha$ array with symmetric tolerances, 64 oscillators, and a coupling strength of $\alpha$ = 0.25 at a tolerance of ±0.1%.

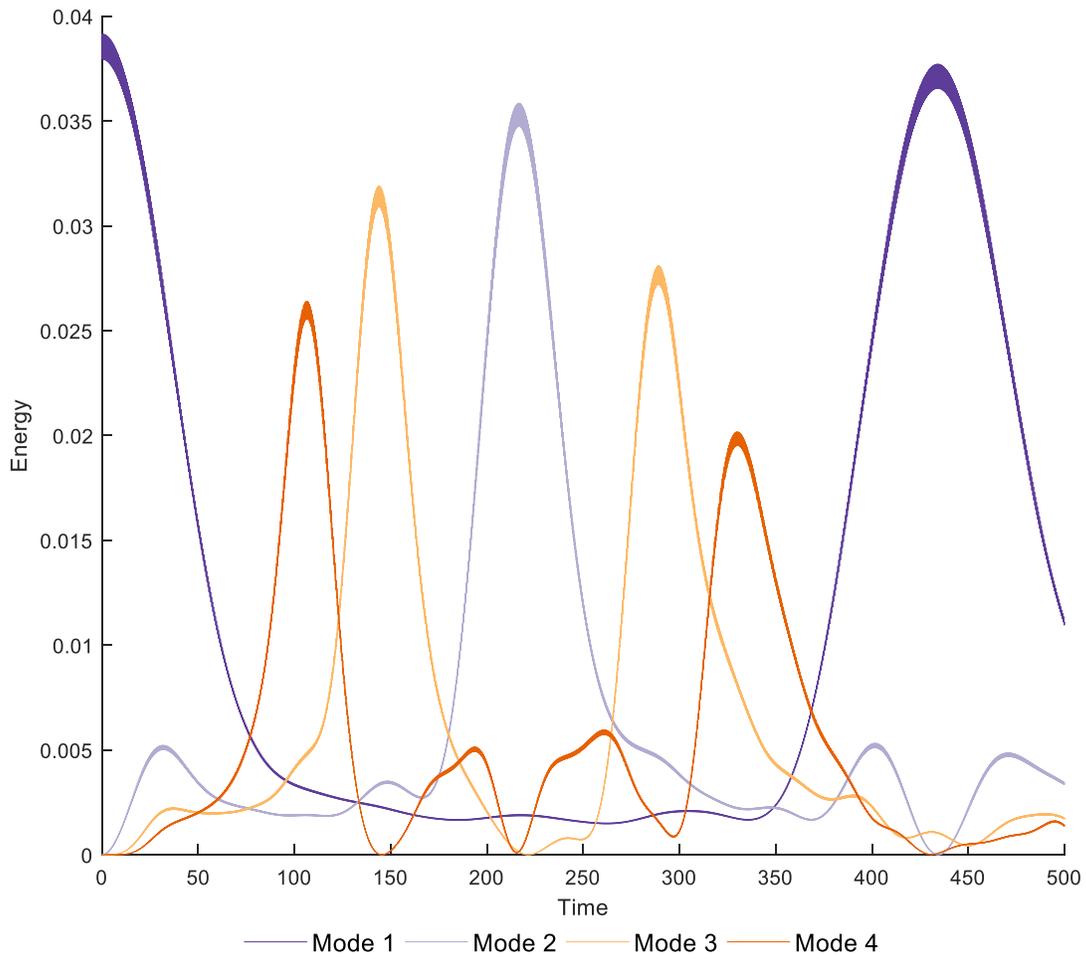

**Fig. S2.**
Simulation of a 1D FPUT-$\alpha$ array with asymmetric tolerances, 64 oscillators, and a coupling strength of $\alpha$ = 0.25 at a tolerance of ±0.1%.